\documentclass[prb,twocolumn,aps,showpacs,fixfloats]{revtex4}
\usepackage{graphicx}
\usepackage{bm}
\usepackage{amsmath,amssymb}
\usepackage{subfigure}
\usepackage{float}
\usepackage{latexsym}
\usepackage{epstopdf}
\usepackage{color}
\usepackage{enumerate}

\begin{document}
\newcommand{\s}{\scriptscriptstyle}
\newcommand{\uu}{\uparrow \uparrow}
\newcommand{\ud}{\uparrow \downarrow}
\newcommand{\du}{\downarrow \uparrow}
\newcommand{\dd}{\downarrow \downarrow}
\newcommand{\ket}[1] { \left|{#1}\right> }
\newcommand{\bra}[1] { \left<{#1}\right| }
\newcommand{\bracket}[2] {\left< \left. {#1} \right| {#2} \right>}
\newcommand{\vc}[1] {\ensuremath {\bm {#1}}}
\newcommand{\tr}{\text{Tr}}

\title{Tunnel magnetoresistance in organic spin valves in the regime of multi-step   tunneling}

\author{R. C. Roundy  and M. E. Raikh} \affiliation{Department of Physics and
Astronomy, University of Utah, Salt Lake City, UT 84112}

\begin{abstract}
A model of a spin valve in which electron transport between the magnetized electrodes
is due to multistep tunneling is analyzed. Motivated by recent experiments on organic spin valves,
we  assume that spin memory loss in the course of transport is due to random hyperfine fields acting on electron while it waits for the next tunneling step. Amazingly, we identify the three-step configurations of sites, for which the tunnel magnetoresistance (TMR) is {\em negative}, suggesting
that the resistance for antiparallel  magnetizations of the electrodes is smaller than for parallel magnetizations.
We analyze
the phase volume of these configurations with respect to magnitudes and relative orientations of the on-site hyperfine fields. The effect of sign reversal of TMR is exclusively due to interference of
the spin-flip amplitudes on each site, it does not emerge within commonly accepted probabilistic description of spin transport. Another feature specific to multistep inelastic tunneling is
{\em bouncing} of electron between nearest neighbors while awaiting a ``hard'' hop. We demonstrate that this bouncing, being absolutely insignificant for conduction of current, can strongly affect the spin memory
loss. This effect is also of interference origin.

\end{abstract}

\maketitle

\section{Introduction.}
A spin valve is a device the resistance of which, $\text{R}_{\s \uu}$ or $\text{R}_{\s \ud}$
depends on the mutual orientation ($\uu$ or $\ud$) of magnetization directions in ferromagnetic electrodes.
Quantitative measure of the effectiveness of a spin valve is the
tunnel magnetoresistance\cite{julliere,review} which is expressed via the
electrode polarizations, $\mathcal{P}_{\s 1}$ and $\mathcal{P}_{\s 2}$, as follows
\begin{equation}
\label{TMR0}
\text{TMR} = \frac{\Delta \text{R}}{\text{R}_{\s \uu}}
 = \frac{\text{R}_{\s \ud} - \text{R}_{\s \uu}}{\text{R}_{\s \uu}}
 = \frac{2\mathcal{P}_{\s 1}\mathcal{P}_{\s 2}}{1-\mathcal{P}_{\s 1}\mathcal{P}_{\s 2}},
\end{equation}
If the thickness, $\text{L}$, of the active layer is large enough,
the spin orientation of injected electrons is ``forgotten'' in course
of transport between the electrodes. Usually this effect is taken
into account by multiplying the product  $\mathcal{P}_{\s 1}\mathcal{P}_{\s 2}$
by a factor $\exp(-\text{L}/l_{\s s})$, where $l_{\s s}$ is the spin-diffusion length.
\begin{figure}[!h]
\includegraphics[width=77mm,clip]{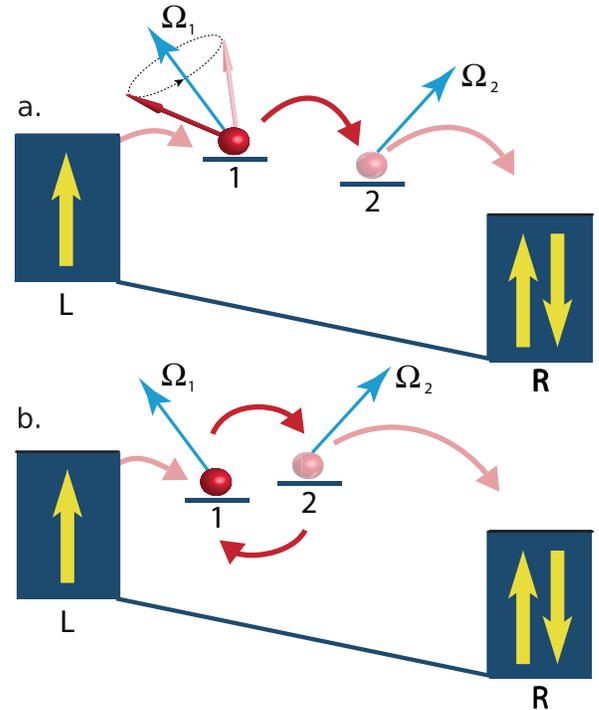}
\caption{(Color online) (a) Illustration of the regime of transport between ferromagnetic electrodes, $L$ and $R$,
dominated by hops via intermediate sites $1$ and $2$. Spin precession in the hyperfine fields takes   place while electron waits for the hops $1\rightarrow 2$ and $2\rightarrow R$. Bias is assumed large, so that all hops are unidirectional; (b) When the sites $1$ and $2$ are close in energy, electron
bounces $2\rightarrow 1\rightarrow 2$ many times while waiting for the ``long'' hop $2\rightarrow R$.}
\label{twosite}
\end{figure}
The use of the concept of spin diffusion implies that, while traveling between the electrodes,  electron experiences many scattering events, and for each event the spin rotation is
weak. Under these conditions the spin polarization is a continuous function of the coordinate.
More generally, the product $\mathcal{P}_{\s 1}\mathcal{P}_{\s 2}$ should be multiplied by $(1-2P_{\s \text{sf}})$, so that
\begin{equation}
\label{TMR1}
\text{TMR} =
\frac{2\mathcal{P}_{\s 1} \mathcal{P}_{\s 2} (1-2P_{\s \text{sf}})}
{1-\mathcal{P}_{\s 1} \mathcal{P}_{\s 2} (1-2P_{\s \text{sf}}) },
\end{equation}
where $P_{\s \text{sf}}$ is the probability that electron flips its spin over the
distance $\text{L}$. Then Eq. (\ref{TMR1}) applies even when the spin rotation in course of
a scattering event is not small, i.e. the initial spin orientation is ``forgotten'' after only a few events.
The factor $(1-2P_{\s \text{sf}})$ emerges in Eq.~(\ref{TMR1}) if one takes into account that, as a result of spin-flips in the active layer,
the states with spin, say, $\uparrow$, in the left electrode are coupled to the states $\uparrow$ in the right electrode with probability $1-P_{\s \text{sf}}$
and to the states  $\downarrow$ with probability $P_{\s \text{sf}}$.
Although Eq. (\ref{TMR1}), for the particular case $(1-2P_{\s \text{sf}})=\exp(-\text{L}/l_{\s s})$ appears in many sources,
for completeness, we present its derivation in the Appendix.

In the present paper we assume that the underlying mechanism responsible for $P_{\s \text{sf}}$
is the spin rotation in hyperfine magnetic fields. This situation is generic for
organic spin valves.\cite{ValveAllMetal, valve1, valve2, valve3, valve4, valve5, valve6, valve7}
In Ref. \onlinecite{Bobbert} experimental data on spin valves with an organic
active layer was analyzed.  The results were interpreted
within a model in which the tunnel transport through the active layer proceeds
in two steps:  first tunneling from the
left electrode $L$ (see Fig. \ref{twosite}) to a localized state in the middle, and,
subsequently, to the right electrode $R$. This ``stop'' near the middle of the active
layer increases the overall tunnel probability from $\exp(-\text{L}/a)$ to $\exp(-\text{L}/2a)$,
where $a$ is the under-barrier tunneling length.
At the same time, while electron waits  to tunnel into $R$, its spin is subject to
a  hyperfine magnetic field created by surrounding  nuclei.
If the average waiting time is $\tau$, the expression for $P_{\s \text{sf}}$ takes the form
\begin{equation}
\label{sf1}
P_{\s \text{sf}}^{(0)} = \frac{1}{2}
\frac{ \left(\Omega^2- \Omega_{\s z}^2\right) \tau^2}{1 + \Omega^2 \tau^2 },
\end{equation}
where $\Omega$ is the total magnetic field at the site (in frequency units),  and $\Omega_{\s z}$
is the projection of this field on the direction of magnetization;
$z$-direction  is determined
by the magnetization in the electrode $L$.

Upon gradual increase of the thickness, the transport will be dominated by three-step tunneling,
then four-step tunneling, and so on.\cite{multisteps}   Rigorous treatment\cite{Tartakovskii} demonstrates that the number of steps, $N$, grows
with the thickness, $\text{L}$, as $N=\sqrt{L/a}$.  In the present paper we
study in detail the domain of lengths
where the transport is via three-step tunneling, as illustrated in Fig. \ref{twosite}. This regime is still analytically tractable, and
yet reveals  fundamental features which are germane to multistep transport and are lacking in the
two-step regime. These features are:

(i) TMR is strongly affected by the fact that the {\em amplitude} for the net spin rotation is
the sum of amplitudes for the rotations  taking place
when electron waits for the hop on site $1$ and on site $2$. We show that this addition of amplitudes
rather than probabilities can lead to {\em negative} TMR, and explore the domain in which the sign reversal of TMR occurs.

(ii) If the waiting time for the hop $2\rightarrow R$ is long, the electron
{\em bounces} between the sites $1$ and $2$ while awaiting  the hop  $2\rightarrow R$.
This bouncing, which has absolutely no effect on the current, can strongly affect the spin rotation.

Both above findings have quantum interference at their core.  In this regard note, that, while electron hops are incoherent,
the spin evolution in course of these hops remains {\em fully coherent}.
The fact that the times spent by electron on each site are random tends to average out the interference effects. It is thus nontrivial that interference effects survive this averaging, and manifest themselves  in the limit $\Omega\tau \gg 1$, when the typical spin rotation is strong.

The paper is organized as follows. In Sect. II we consider the transport via two sites at
 high bias when electron moves only forward. In Sect. III. we relax this condition and allow
fast backward hops while awaiting the slow forward hop. For both situations we calculate $P_{\s \text{sf}}$ averaged over the random durations of the waiting periods, which should be substituted into Eq. (\ref{TMR1}). We pay special attention to $P_{\s \text{sf}}$ in the presence of external magnetic field in view of mysterious absence of the Hanle effect in spin
valves reported recently\cite{noHanle1,noHanle2}.
In Sect. IV we discuss the implications of our findings for true multistep or bulk transport.

\section{Interference correction to the two-step spin-flip probability}
\subsection{Analytical expression for $P_{\s \text{sf}}$.}

Under a strong applied bias the motion of the electron is  unidirectional.
The hops proceed in a sequence  $L\rightarrow 1\rightarrow 2\rightarrow R$.
Denote with $t_{\s 1}$ and $t_{\s 2}$ the random times spent by electron on sites $1$ and $2$,
respectively. The evolution of spin is described by the product of the unitary matrices
$U(t_{\s 2})U(t_{\s 1})$, where the matrix $U(t)$ is defined as
\begin{equation}
\label{time-evolution}
U(t)  =
 \begin{bmatrix}
 \cos\alpha - i \frac{\Omega_{\s z}}{\Omega}
 \sin\alpha
 &
-i \frac{\Omega_{\s -}}{\Omega} \sin \alpha
 \\
-i \frac{\Omega_{\s +}}{\Omega} \sin \alpha
&
 \cos\alpha + i \frac{\Omega_{\s z}}{\Omega}
 \sin\alpha
\end{bmatrix},\; \alpha=\frac{\Omega t}{2},
\end{equation}
where $\Omega_{\s \pm}=\Omega_{\s x} \pm i\Omega_{\s y}$.
The  spin-flip amplitude  is given by a non-diagonal element, $A_{\s \ud} = -i \frac{\Omega_{\s +}}{\Omega} \sin \left( \frac{\Omega t}{2} \right)$. Averaging of $p_{\s \ud} = |A_{\s \ud}|^2$ over the Poisson distribution, $\frac{1}{\tau}\exp(-t/\tau)$, of the waiting time,
$t$, reproduces Eq. (\ref{sf1}).

The spin-flip amplitude after two steps is given by non-diagonal element of
$U(t_{\s 2})U(t_{\s 1})$. It can be written in the form
\begin{equation}
{\tilde A}_{\s \ud} = A^{(1)}_{\s \ud} A^{(2)}_{\s \dd} + A^{(1)}_{\s \uu} A^{(2)}_{\s \ud},
\end{equation}
where $A^{(1,2)}$ are the corresponding elements of the matrices $U(t_{\s 1})$ and $U(t_{\s 2})$.
Averaging of $P_{\s \text{sf}}=|{\tilde A}_{\s \ud}|^2$ over random times $t_{\s 1}$, $t_{\s 2}$ can be easily carried out.
First, it is convenient to present $P_{\s \text{sf}}$ in the form
\begin{equation}
\label{partition}
P_{\s \text{sf}}=P_{\s \text{incoh}}+\delta P_{\s \text{int}}
\end{equation}
of the sum of incoherent and interference contributions defined as

\begin{equation}
\label{classical}
P_{\s \text{incoh}} = p^{(1)}_{\s \text{sf}} \left(1- p^{(2)}_{\s \text{sf}}\right) +
\left(1-p^{(1)}_{\s \text{sf}}\right)
p^{(2)}_{\s \text{sf}},
\end{equation}
where $p^{(1)}_{\s \text{sf}}$ and $p^{(2)}_{\s \text{sf}}$ are the partial probabilities given
by Eq. (\ref{sf1}), and
\begin{equation}
\delta P_{\s \text{int}}
 = 2 \langle\text{Re}\left(
A^{(1)}_{\s \ud} A^{(2)}_{\s \dd} A^{(1)*}_{\s \uu} A^{(2) *}_{\s \ud}
\right)\rangle_{\s t_{\s 1},t_{\s 2} }.
\end{equation}
Averaging of $\delta P_{\s \text{int}}$ over $t_{\s 1}$ and $t_{\s 2}$ can be performed independently.
The product of the terms depending on $t_{\s 1}$ is
\begin{multline}
\label{interference}
A^{(1)}_{\s \ud} A^{(1) *}_{\s \uu}
 = \left[ -i \frac{\Omega_{\s 1+}}{\Omega_{\s 1}}\sin\left(\frac{\Omega_{\s 1} t_{\s 1}}{2}\right)
 \right] \\  \times \left[
	\cos\left(\frac{\Omega_{\s 1} t_{\s 1}}{2}\right)
+i \frac{\Omega_{\s 1z}}{\Omega_{\s 1}}\sin\left(\frac{\Omega_{\s 1} t_{\s 1}}{2}\right)
 \right].
\end{multline}
Denote with $\tau_{\s 1}$ the average waiting time for the hop $1\rightarrow 2$.
Averaging of Eq. (\ref{interference}) over $t_{\s 1}$ yields a compact
expression
\begin{equation}
\langle A^{(1)}_{\s \ud} A^{(1) *}_{\s \uu}\rangle_{t_{\s 1}}
 = \frac{1}{2} \frac{  \Omega_{\s 1+} \tau_{\s 1}
 ( -i + \Omega_{\s 1z} \tau_{\s 1} )}
 { 1+ \Omega_{\s 1}^2 \tau_{\s 1}^2 }.
\end{equation}
The same expression with $\tau_{\s 2}$ instead of $\tau_{\s 1}$ and $\Omega_{\s 2}$ instead
of $\Omega_{\s 1}$ together with an additional complex conjugation describes the result of averaging over $t_{\s 2}$.
Altogether, the expression for $\delta P_{\s \text{int}}$ can be cast in the form
\begin{equation}
\label{interference1}
\delta P_{\s \text{int}}
 = \frac{1}{2} \text{Re}\left(
 \frac{\Omega_{\s 1+}\Omega_{\s 2-} \tau_{\s 1} \tau_{\s 2}
 (1 + i\Omega_{\s 1z}\tau_{\s 1})(1 - i\Omega_{\s 2z} \tau_{\s 2})
 }{(1+\Omega_{\s 1}^2\tau_{\s 1}^2)(1+\Omega_{\s 2}^2\tau_{\s 2}^2)}
\right).
\end{equation}
At this point note that, within the probabilistic approach, the result for $P_{\s \text{sf}}$
would be simply $P_{\s \text{incoh}}$.
Indeed,
within this approach, the net spin flip corresponds to
flipping on the first site and preserving spin on the second site
or vice versa.
Since these are  mutually exclusive events their probabilities
simply add.
Because of this,  $\delta P_{\s \text{int}}=P_{\s \text{sf}}-P_{\s \text{incoh}}$ is a measure of quantum interference of the amplitudes of two rotations that took place at site $1$ and at site $2$.

Throughout this subsection we implicitly identified $P_{\s \text{sf}}$ with the spin-flip probability which appears in Eq. (\ref{TMR1}). It is however not entirely obvious that the quantum-mechanical quantity  $P_{\s \text{sf}}(t_{\s 1},t_{\s 2})$ averaged over the Poisson distribution of the waiting times is the same quantity which appears in  Eq. (\ref{TMR1}). Formal
justification is presented in the Appendix.

In the next subsection we analyze several particular cases when the interference term has
dramatic consequences for TMR.

\subsection{Limiting cases}

It is instructive to express
 the result Eq. (\ref{interference})  via the partial probabilities $p_{\s \text{sf}}^{(1)}$ and $p_{\s \text{sf}}^{(2)}$ as
follows
\begin{equation}
\label{expressed}
\delta P_{\s \text{int}}=
\sqrt{ p^{(1)}_{\s \text{sf}} \left( 1-2p^{(1)}_{\s \text{sf}} \right)
p^{(2)}_{\s \text{sf}} \left(1-2p^{(2)}_{\s \text{sf}} \right) }  \cos \phi,
\end{equation}
where the phase $\phi$ is defined as
\begin{equation}
\label{phase}
\phi=\varphi_{\s 1} - \varphi_{\s 2}  + \tan^{-1}(\Omega_{\s 1} \tau_{\s 1} \cos \vartheta_{\s 1})
- \tan^{-1}(\Omega_{\s 2} \tau_{\s 2} \cos \vartheta_{\s 2}).
\end{equation}
The angles $\vartheta_{\s 1}$,  $\varphi_{\s 1}$ $(\vartheta_{\s 2}, \varphi_{\s 2})$ are the  spherical
angles describing the polar and azimuthal orientations of the vector ${\bm \Omega}_{\s 1}$ $({\bm \Omega}_{\s 2})$.
Eqs. (\ref{expressed}), (\ref{phase}) indicate that interference can be either constructive
of destructive depending on the mutual orientations of the fields ${\bm \Omega}_{\s 1}$, ${\bm \Omega}_{\s 2}$.
When $\Omega_{\s 1}\tau_{\s 1}$ and  $\Omega_{\s 2}\tau_{\s 2}$
are of the same order, the interference correction is of  the order of $P_{\s
\text{incoh}}$.

\subsubsection{Identical fields, $p^{(1)}_{\s sf}=p^{(2)}_{\s sf}$}

The role of interference is maximal when the vectors ${\bm \Omega}_{\s 1}$ and ${\bm \Omega}_{\s 2}$
are collinear and $\Omega_{\s 1}\tau_{\s 1} = \Omega_{\s 2} \tau_{\s 2}$.
Then we have

\begin{equation}
\label{quadratic}
P_{\s \text{sf}}= 2p_{\s \text{sf}}(1 - p_{\s \text{sf}}) +
p_{\s \text{sf}} (1-2p_{\s \text{sf}})
 = 3 p_{\s \text{sf}} - 4 p_{\s \text{sf}}^2.
\end{equation}
\begin{figure}[t]
\includegraphics[width=57mm, clip]{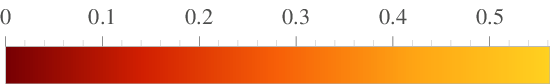}
\includegraphics[width=77mm, clip]{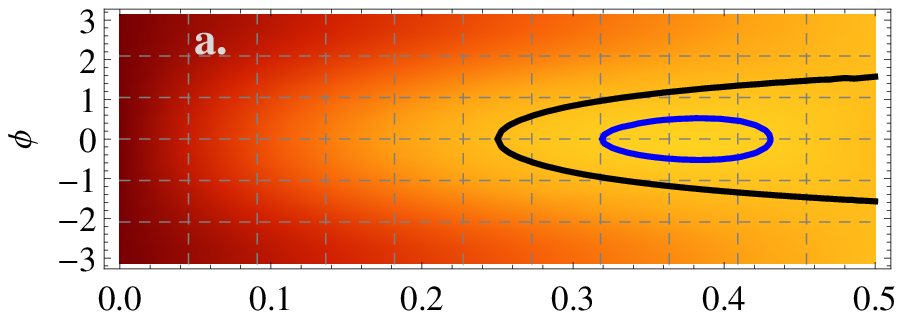}
\includegraphics[width=77mm, clip]{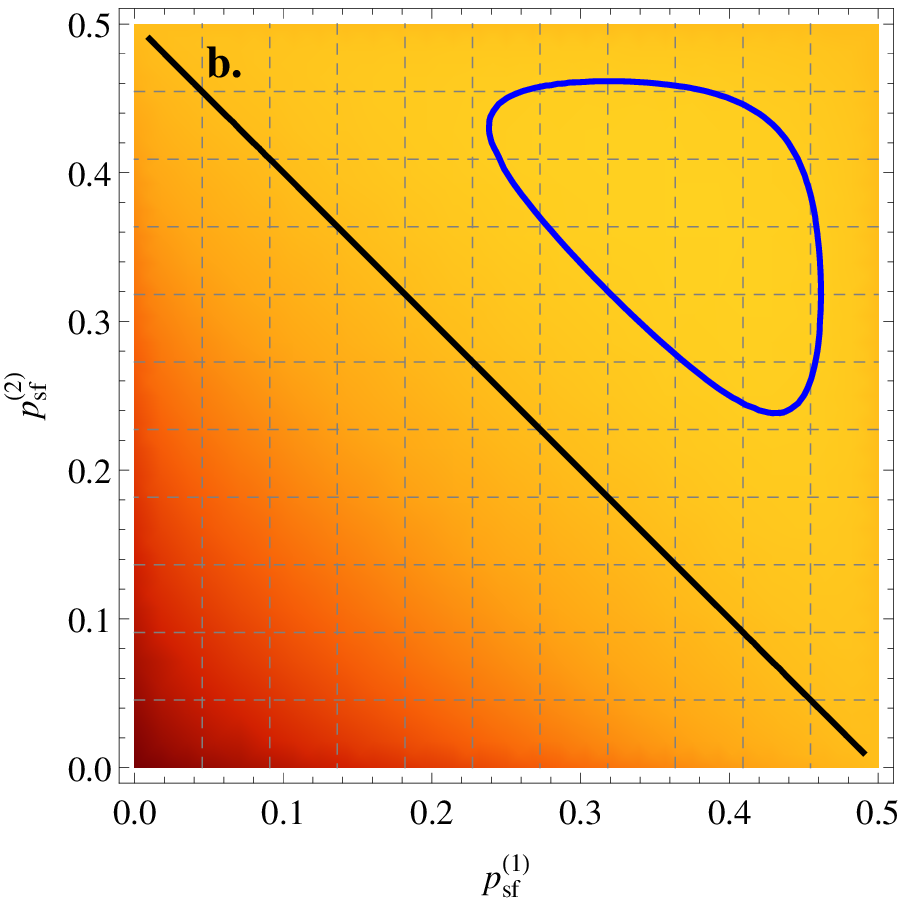}
\caption{(Color online). (a) Contour plot of the cumulative spin-flip probability, $P_{\s \text{sf}}$, calculated from Eqs. (\ref{classical}) and (\ref{expressed}). It is assumed that partial probabilities, $p_{\s \text{sf}}^{(1)}$ and $p_{\s \text{sf}}^{(2)}$,  are the same (horizontal axis), while
the hyperfine fields at sites $1$ and $2$ are skewed by angle $\phi$ (vertical axis).
Black curve separates the domains with positive TMR (to the left) and negative TMR (to the right).  Blue curve is a contour $P_{\s \text{sf}}=0.55$. (b) Same as (a) for the limiting case
when hyperfine fields are parallel, $\phi=0$, but have different magnitudes, so that the partial probabilities $p_{\s \text{sf}}^{(1)}$ )(horizontal axis) and $p_{\s \text{sf}}^{(2)}$ (vertical axis) are different. As $\phi$ increases, the domain of negative TMR shrinks and completely disappears at $\phi=\pi/2$. }
\label{allowed}
\end{figure}
To illuminate the non-triviality of
Eq. (\ref{quadratic}), note that the single-scattering value, $p_{\s \text{sf}}$, never exceeds $1/2$.  Equally the incoherent part of the two-scattering probability, $P_{\s \text{incoh}}$,
never exceeds $1/2$.  The physical meaning of these restrictions is obvious: $p_{\s \text{sf}} = 1/2$
implies a full loss of the spin memory. Therefore, if  {\em either}  of two values of $p_{\s \text{sf}}$
in Eq. (\ref{classical}) is equal to $1/2$, we get $P_{\s \text{incoh}}=1/2$ regardless
of the value of the other $p_{\s \text{sf}}$. Interestingly, the exact $P_{\s \text{sf}}$ does not satisfy
this restriction. Similarly to $P_{\s \text{incoh}}$,   Eq. (\ref{quadratic}) does yield $1/2$, for $p_{\s \text{sf}}=1/2$, when the interference term vanishes.
However, the value of
$P_{\s \text{sf}}$ can actually {\em exceed} $1/2$ for smaller $p_{\s \text{sf}}$. Namely, at $p_{\s \text{sf}} = 3/8$,
Eq. (\ref{quadratic}) has a maximum and assumes the value $P_{\s \text{sf}} = 9/16$.
This implies that the TMR, defined by Eq. (\ref{TMR0}), is {\em negative} for this $p_{\s \text{sf}}$.  Moreover, it retains negative value within the domain $1/4 < p_{\s \text{sf}}< 1/2$.
Physically, this means that the resistance for antiparallel orientations of magnetization
in the electrodes is {\em smaller} than for the parallel orientation.

In fact, negative values of TMR happen not only when
the vectors
${\bm \Omega_{\s 1}}$ and   ${\bm \Omega_{\s 2}}$ coincide.
For illustration, assume that the product $\Omega_{\s 1}\tau_{\s 1}$ is still
equal to  $\Omega_{\s 2}\tau_{\s 2}$, but the vectors ${\bm \Omega_{\s 1}}$ and   ${\bm \Omega_{\s 2}}$ are
skewed by an angle $\phi$. The domain $P_{\s \text{sf}}=1/2$  on the $\left(p_{\s \text{sf}}, \phi\right)$-plane is shown in Fig. \ref{allowed}. The ``allowed'' values of $\phi$ range from $0$ at $p_{\s \text{sf}}=1/4$
to $\pm \pi/2$ at $p_{\s \text{sf}}=1/2$.

To what degree is the assumption that the field magnitudes are precisely
equal to each other crucial for negative TMR?  To  answer this question we have plotted
in Fig. \ref{allowed}b, the contour plot of $P_{\s \text{sf}}$ for configurations with $\phi = 0$
when $p^{(1)}_{\s \text{sf}}$ and $p^{(2)}_{\s \text{sf}}$ vary over their
allowed values. We see that negative TMR corresponds to the domain above the diagonal of
the square. This domain shrinks upon increasing $\phi$.

\subsubsection{Identical fields, many hops}

\begin{figure}
\includegraphics[width=77mm,clip]{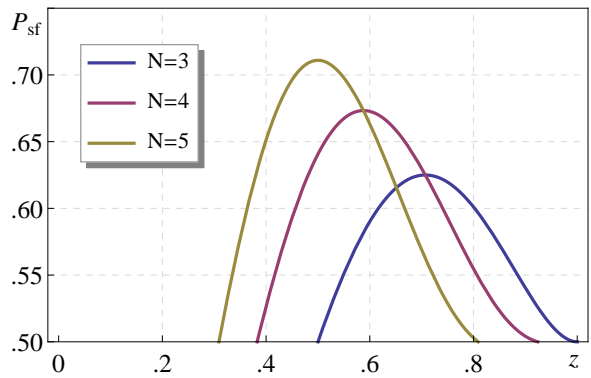}
\caption{(Color online) The spin-flip probability for $N$-step process is plotted from Eq. (\ref {NP_sf}) versus dimensionless combination $z=\Omega\tau/(1+\Omega^2\tau^2)^{1/2}$ for $N=3,4,5$. It is assumed that in-plane hyperfine fields $\Omega$, and waiting times, $\tau$, are the same at all $(N-1)$ sites. Only the parts of the curves for which TMR is negative are shown.}
\label{chebyshev}
\end{figure}

In the example considered above the TMR was ``most negative''  when both hyperfine fields were equal, i.e. the hopping of electron does not interrupt the spin precession at all.
It might seem
that this case should be reducible to the precession in one given field for which the result
Eq. (\ref{sf1})  never goes above $1/2$.  The resolution lies in the fact that
 Eq. (\ref{sf1}) was obtained upon averaging over exponential distribution
of the waiting times.  When {\em two} hops are performed in the
same magnetic field, the distribution function of the two-hop waiting times
is different: $F_{\s 2}(T) = T/\tau^2 \exp(-T/\tau)$.  It is because of
this difference that $P_{\s \text{sf}} > 1/2$ emerges.  In this regard,
it is interesting to consider what  happens if an electron
performed $N>2$ steps in the same magnetic field.  Then the distribution
function of the waiting time is
\begin{equation}
\label{distribution}
F_{\s N}(T) = \frac{T^{N-1}}{\tau^{N}(N-1)!} \exp(-T/\tau).
\end{equation}
With this distribution, the expression for spin-flip probability can be easily shown to take
the form
\begin{equation}
\label{NP_sf}
P_{\s \text{sf}} = \frac{|\Omega_{\s +}|^2}{2 \Omega^2}
\left[
    1 - \frac
            {\cos\left( N \sin^{-1}
                \frac{\Omega \tau}
                    {\sqrt{1 + \Omega^2 \tau^2}} \right)}
        {\left( 1+ \Omega^2 \tau^2 \right)^{N/2} }
 \right].
\end{equation}
The situation most favorable for negative TMR is an in-plane orientation of magnetic field, when the prefactor
in Eq. (\ref{NP_sf}) is equal to $1/2$. Then we have $P_{\s \text{sf}}>1/2$ in the domains of $\Omega\tau$ when the cosine is negative. These domains are shown in Fig. \ref{chebyshev} for several values of $N$. We see that the
net width of the domains with negative TMR does not 
change much with $N$, while the
magnitude of negative TMR 
grows with increasing $N$.

Another message of Eq.(\ref{NP_sf})
is that  $P_{\s \text{sf}}$ saturates with damped oscillations upon
increasing $N$.  The saturation value is $|\Omega_{\s +}|^2/2\Omega^2$.
This saturation is the result of quantum interference.  To illuminate
this point, let us compare it to the result obtained via probabilistic
treatment, i.e. neglecting interference
\begin{equation}
\label{NP_sf-prob}
P_{\s \text{sf}} = \frac{1}{2}\left(
1 - \exp\left\{- N \left| \ln \left(
1 - \frac{|\Omega_{\s +}|^2\tau^2}{1+\Omega^2 \tau^2}
\right) \right| \right\}
\right).
\end{equation}
We see that neglecting quantum evolution leads to the intuitively obvious
prediction that in the limit of large $N$, $P_{\s \text{sf}}$ approaches $1/2$ exponentially.
The logarithm in the exponent relates the ``time'' of spin-memory loss to the
hyperfine field magnitude.\cite{Bobbert2}
The dramatic difference between Eq. (\ref{NP_sf}) and Eq. (\ref{NP_sf-prob})
indicates that interference survives in spite of the fact that the
individual hopping times are random.  The fact that rotation of spin in
a constant magnetic field can be non-trivial due to the randomness in
the waiting times for subsequent hops was previously pointed out in Ref.
 \onlinecite{Flatte1}.

%

\subsubsection{Averaging over hyperfine fields}

It is apparent from Eq. (\ref{expressed}) that, since  $\cos\phi$ is zero on average,
the interference correction to $P_{\s \text{sf}}$ vanishes
upon averaging over hyperfine field distribution.
This explains why the D'yakonov-Perel result\cite{Dyakonov} for the spin relaxation time derived from probabilistic treatment remains valid in spite of the fact that spin rotations for subsequent electron steps are
strongly correlated; large number of electron collisions each of which is accompanied by a small spin rotation\cite{Dyakonov},  guarantees that the averaging takes place.
Equally, the averaging
happens for a spin valve with large area of the active layer.
For a given path through the layer $\delta P_{\s \text{int}}$
can be of the order of $P_{\s \text{sf}}$, but it will not
contribute to the average spin-flip probability coming from
many channels. If the area is finite, so that the
number of channels, $N\gg 1$, is also finite,  the averaging
will be incomplete. The TMR will acquire a random correction
of the order $\Delta /\sqrt{N}$, where $\Delta^2$ is the variance
of $P_{\s \text{sf}}$, which we calculate below.

It follows from Eqs. (\ref{partition}), (\ref{expressed})  that the variance has two contributions

\begin{equation}
\label{Delta}
\Delta^2=
\overline{P_{\s \text{sf}}^2}-\Bigl(\overline{P_{\s \text{sf}}}\Bigr)^2 =\Delta_{\s \text{incoh}}^2+\Delta_{\s \text{int}}^2.
\end{equation}
where $\Delta_{\s \text{incoh}}^2$ and  $\Delta_{\s \text{int}}^2$ are the
variances of the incoherent and coherent contributions, respectively. The
overline stands for hyperfine averaging over the gaussian distribution,
$\frac{1}{\sqrt{\pi}b_{\s 0}}\exp[-b_{\s i}^2/b_{\s 0}^2]$, of the
hyperfine-field components, $b_{\s i}$.  Then the variance
$\Delta_{\s \text{incoh}}^2$ can be expressed through averages
$\overline{p}_{\s 1,2}$ and partial variances
$\Delta_{\s 1,2}^2 = \overline{p^2}_{\s 1,2} -
 \left( \overline{p}_{\s 1,2} \right)^2$ as follows
\begin{equation}
\label{variance-incoh}
\Delta_{\s \text{incoh}}^2 =
 (1-2\overline{p}_{\s 2})^2\Delta_{\s 1}^2
+ (1-2\overline{p}_{\s 1})^2 \Delta_{\s 2}^2 + 4 \Delta_{\s 1}^2\Delta_{\s 2}^2.
\end{equation}
The corresponding expression for the interference contribution, $\Delta_{\s \text{int}}^2$, reads
\begin{multline}
\label{variance-int}
\Delta_{\s \text{int}}^2 =
\frac{1}{2}\left[ (\overline{p}_{\s 1} + 2 \overline{p}_{\s 1}^2)(\overline{p}_{\s 2} + 2 \overline{p}_{\s 2}^2)
- 2  (\overline{p}_{\s 2} + 2 \overline{p}_{\s 2}^2)\Delta_{\s 1}^2 \right. \\
 \left. - 2 (\overline{p}_{\s 1} + 2 \overline{p}_{\s 1}^2)
\Delta_{\s 2}^2 + 4 \Delta_{\s 1}^2 \Delta_{\s 2}^2
\right].
\end{multline}
Analytical expressions for $\overline{p}_{\s 1,2}$ and $\Delta^2_{\s 1,2}$ take a simple form
in the limits of strong ($\Omega\tau \gg 1$) and weak ($\Omega\tau \ll 1$) magnetic fields:

\begin{equation}
\label{p-bar}
\overline{p}_{\s 1,2} = \left\{ \begin{array}{ll}
b_{\s 0}^2 \tau_{\s 1,2}^2,  & \Omega_{\s 1,2} \tau_{\s 1, 2} \ll 1 \\
\frac{1}{2}\int\limits_0^\infty \frac{ds}{(1+s)^{5/2}}
\exp\left[ -\frac{s}{1+s} \frac{B^2}{b_{\s 0}^2} \right],
 & \Omega_{\s 1,2} \tau_{\s 1, 2} \gg 1
\end{array} \right.
\end{equation}

\begin{equation}
\label{Delta-final}
\Delta_{\s 1,2}^2 = \left\{ \begin{array}{ll}
b_{\s 0}^4 \tau_{\s 1,2}^4,  & \Omega_{\s 1,2} \tau_{\s 1, 2} \ll 1 \\
-\overline{p}_{\s 1,2}^2 + \frac{1}{2}\int\limits_0^\infty  \frac{s \, ds }{(1+s)^{7/2}}
\exp\left[ -\frac{s}{1+s} \frac{B^2}{b_{\s 0}^2} \right],
 & \Omega_{\s 1,2} \tau_{\s 1, 2} \gg 1
\end{array} \right.
\end{equation}
Here $B$ is the external field directed along the $z$-axis.
Eq. (\ref{p-bar}) describes the fall-off of the disorder-averaged spin-flip probability with $B$. For weak hyperfine field, $b_{\s 0}\tau \ll 1$, the dependence  $\overline{p}_{\s 1,2}(B)$
evolves from small value, $b_{\s 0}^2\tau_{\s 1,2}^2$, to $b_{\s 0}^2/2B^2$.
In the opposite limit,  $b_{\s 0}\tau \gg 1$, the evolution starts from $\overline{p}_{\s 1,2}(0)=1/3$ and converges to $b_{\s 0}^2/2B^2$ when $B$ exceeds $b_{\s 0}$.

In the first case we have $\Delta_{\s \text{int}}^2\approx \overline{p}_{\s 1} \overline{p}_{\s 2} / 2$, while
$\Delta_{\s \text{incoh}}^2\approx \Delta_{\s 1}^2 + \Delta_{\s 2}^2$, so
that for the ratio  $\Delta_{\s \text{int}}^2/\Delta_{\s \text{incoh}}^2$ we get $\tau_{\s 1}^2\tau_{\s 2}^2/\left(\tau_{\s 1}^4+\tau_{\s 2}^4\right)$,
i.e. the interference contribution
is of the same order as $\Delta_{\s \text{incoh}}^2$.
For strong hyperfine field,
$\Delta_{\s \text{int}}^2$
and $\Delta_{\s \text{incoh}}^2$ do not depend on $\tau$.
At $B=0$ Eq. (\ref{Delta-final}) yields $\Delta_{\s 1,2}^2=\frac{1}{45}$. Using this value, we get for the contributions to the variance:
$\Delta_{\s \text{incoh}}^2=\frac{7}{2}\left(\frac{2}{45}\right)^2$ and
$\Delta_{\s \text{int}}^2=\frac{1}{2}\left(\frac{23}{45}\right)^2$, i.e. the
interference contribution is almost $20$ times bigger than the incoherent contribution.
Finally, consider the limit of strong hyperfine field and $B\gg b_{\s 0}$. In this limit Eq. (\ref{Delta-final}) yields $\Delta_{\s 1,2}^2=b_{\s 0}^4/2B^4$, and we thus have:
\begin{equation}
\Delta_{\s \text{int}}^2= \frac{\overline{p}_{\s 1,2}^2}{2}=\frac{b_{\s 0}^4}{8B^4}, \quad \Delta_{\s \text{incoh}}^2=2\Delta_{\s 1,2}^2=\frac{b_{\s 0}^4}{B^4}=8\Delta_{\s \text{int}}^2.
\end{equation}

In summary, for all the domains of change of the dimensionless parameters $b_{\s 0}\tau$ and $b_{\s 0}/B$ the variance, $\Delta$, of the
spin-flip probability is of the order of average $P_{\s \text{sf}}$, and the interference  contribution to $\Delta$ is
comparable to $\Delta$ itself.

In conclusion of the Section, note that for $\tau_{\s 2} \gg \tau_{\s 1}$ the hops $1 \rightarrow 2$ between the sites do not affect the current. Except for anomalous configurations of hyperfine fields, when $\Omega_{\s 2\perp}$ is much smaller than $\Omega_{\s 1\perp}$,
 these hops
also do not affect the spin memory. In the next Section we will demonstrate that multiple {\em bounces}
of electron within a pair of sites,
while not affecting the current,
can significantly affect the spin memory. This effect, caused by interference, is most pronounced in the presence of an external magnetic field.

The partial spin-flip probabilities obviously fall off with
magnetic field, $B$, which is parallel to the polarization in the injector. The result of the probabilistic approach,   $P_{\s \text{incoh}}$, also falls off with  $B$. As it is easy to see from Eq. (\ref{sf1}), the probability $p_{\s \text{sf}}$
is proportional to $1/B^2$ for $\Omega\tau  \gg 1$.
 Concerning the magnitude of the interference term,
Eq. (\ref{interference}), it can actually grow with $B$ if both partial probabilities, $p_{\s \text{sf}}$, exceed $1/4$. However, when they are both small, the magnitude of interference term also drops with $B$ as $1/B^2$.  In the next Section we will demonstrate that electron bounces can transform the
$1/B^2$ to a much weaker dependence.

\section{Effect of bouncing on the spin-flip probability}

Assume that $\tau_{\s 2}$ is much bigger than $\tau_{\s 1}$ and the activation
energy for the back-hop $2\rightarrow 1$ is small, Fig. \ref{twosite}b. In this case, as
it was explained in the Introduction, while awaiting the hop $2\rightarrow R$,
the electron performs $m=\tau_{\s 2}/\tau_{\s 1}\gg 1$ hops $2\rightarrow 1$ and back.
This bouncing affects strongly the spin-rotation and enhances the interference
contribution to $P_{\s \text{sf}}$.

Note first that, within the probabilistic description,  taking bounces into account
is equivalent to modifying the partial probability $p^{(1)}_{\s \text{sf}} $
 \begin{equation}
 \label{probabilistic}
{\tilde p}^{(1)}_{\s \text{sf}}= \frac{1}{2} - \frac{1}{2}(1-2p_{\s \text{sf}})^m,
\end{equation}
where $m$ is odd. Eq. (\ref {probabilistic}) expresses the fact that  ${\tilde p}^{(1)}_{\s \text{sf}}$ is the sum of probabilities to flip spin {\em only} once in the course of all bounces,
only three times in the course of all bounces, and so on.
Accumulation of the  powers of $(1-2p_{\s \text{sf}})$ with $m$ is natural since $(1-2p_{\s \text{sf}})$ is the probability of spin preservation for one step.
In reality, while bouncing,
electron spin experiences an {\em alternating} magnetic field,
which takes only two values. This favors the interference processes, and the result Eq. (\ref{probabilistic}) should be
compared to the result of treatment with interference taken into
account. Within the latter treatment, the spin-flip amplitude
is given by the non-diagonal elements of the matrix product
$U_{\s m}(t_{\s m})...U_{\s 2}(t_{\s 2})U_{\s 1}(t_{\s 1})$, where $U_{\s j}(t_{\s i})$ is the matrix Eq. (\ref{time-evolution}) in
which the fields corresponding to $U_{\s 1}$   and
$U_{\s 2}$ are ${\bm \Omega}_{\s 1}$ and ${\bm \Omega}_{\s 2}$, respectively. The times,  $t_{\s i}$, are random, but have the same distribution.

To illuminate the importance of interference in course of bouncing, consider the following simple example. Suppose that
$m=3$ and that the external field is strong, i.e. $\Omega_{\s \perp}\ll \Omega$. Assume as well, that the in-plane field components for all three steps are equal in magnitude and differ
only in azimuthal  orientations, $\chi_{\s i}$.
Then the non-diagonal matrix element of the  product

\vspace{6pt}
\resizebox{\linewidth}{!}{
$\begin{pmatrix}
u & -i v e^{i \chi_{\s 3}} \\
-i v e^{-i \chi_{\s 3}} & u
\end{pmatrix}
\begin{pmatrix}
u & -i v e^{i \chi_{\s 2}} \\
-i v e^{-i \chi_{\s 2}} & u
\end{pmatrix}
\begin{pmatrix}
u & -i v e^{i \chi_{\s 1}} \\
-i v e^{-i \chi_{\s 1}} & u
\end{pmatrix}
$}

\vspace{6pt}
\noindent takes a simple form
\begin{equation}
\label{A}
\tilde{A} = iv^3e^{-i(\chi_{\s 3} - \chi_{\s 2} + \chi_{\s 1})}
 - iu^2v\left( e^{-i \chi_{\s 1}} + e^{-i \chi_{\s 2}} + e^{-i \chi_{\s 3}} \right).
\end{equation}
Here $u^2+v^2=1$.
For a sequential hopping all $\chi_{\s i}$ are random. Then the average value of $\vert \tilde{A}\vert^2$ is given by
\begin{equation}
\label{average1}
|\tilde{A}|^2 = v^6 + 3u^4v^2.
\end{equation}
On the other hand, if the hops constitute a single bounce
$1\rightarrow 2 \rightarrow 1$, we have $\chi_{\s 1}=\chi_{\s 3}$, which
leads to the following expression for average $\vert \tilde{A}\vert^2$.
\begin{equation}
\label{average2}
|\tilde{A}|^2 = v^6 + 5 u^4v^2.
\end{equation}
The result Eq. (\ref{average1}) can be brought in correspondence with probabilistic description Eq. (\ref{probabilistic}), if we
identify $\vert v\vert^2$ with $p_{\s \text{sf}}$.
The fact that Eq. (\ref{average2}) yields a bigger value for $\vert \tilde{A}\vert^2$ is due to interference of the spin-flip amplitudes
which arises as a result of visiting the site $1$ {\em twice}.
Multiple bouncing would amplify the role of interference.
It is easier to capture this effect quantitatively by starting directly from the Schr\"{o}dinger equation for electron spin in
a time-dependent magnetic field.

In the next two subsections we will separately consider the effect of bouncing on
the spin preservation in a zero and in strong external fields.
We will demonstrate that  in these two limits the effects of bouncing are {\em opposite}.

\subsection{Bouncing in a zero external field}

The amplitudes  $a_{\s 1}$ and $a_{\s 2}$
for  an electron to have an $\uparrow$ and  $\downarrow$
projections of spin satisfy the system
\begin{align}
\label{system}
i \dot{a}_{\s 1}(t) &= \frac{1}{2}
	\Bigl[ b_{\s z}(t) a_{\s 1}(t)  + b_{\s \perp}^{*}(t) a_{\s 2}(t) \Bigr], \nonumber\\
i \dot{a}_{\s 2}(t) &= \frac{1}{2}
	\Bigl[ b_{\s \perp}(t) a_{\s 1}(t)  - b_{\s z}(t) a_{\s 2}(t) \Bigr].
\end{align}
Suppose that at time $t=0$ electron spin is directed $\uparrow$,
so that $a_{\s 2}=0$. A formal solution of the system Eq. (\ref{system})
reads
\begin{equation}
\label{formal}
a_{\s 2}(t) = \frac{-i}{2} \int\limits_0^t dt' \,
b_{\s \perp}(t') a_{\s 1}(t') \exp\left[ \frac{-i}{2} \int\limits_0^{t'} dt'' \, b_{\s z}(t'') \right].
\end{equation}
If the net spin rotation during the time, $\tau_{\s 2}$, when electron waits for the hop $2 \rightarrow R$ is small, we can set $a_{\s 1}(t)=1$ and $\exp[-\int_0^t b_{\s z}(t^{\prime})] =1$
in the integrand. This leads to the following result for the spin-flip probability
\begin{equation}
\label{discrete}
P_{\s \text{sf}} =  \left|
b_{\s 1 \perp} \left( t_{\s 1} + t_{\s 3} + \cdots  \right)
+
b_{\s 2 \perp} \left( t_{\s 2} + t_{\s 4} + \cdots  \right)
\right|^2,
\end{equation}
where $t_{\s 1}, t_{\s 3}, \ldots$ are the time intervals spent by electron on the site $1$,
while $t_{\s 2}, t_{\s 4}, \ldots$ are the time intervals spent by electron on the site $2$; each time interval is $\sim \tau_{\s 1}$. It is an important consequence of bouncing that these time
intervals add up, instead of averaging out, which would be the case for hopping
 over multiple sites. The big parameter $\tau_{\s 2}/\tau_{\s 1}$ allows us to replace these sums by $\tau_{\s 2}/2$. Then we get
\begin{equation}
\label{averaging}
P_{\s \text{sf}} =  \frac{\tau_{\s 2}^2}{2} \left|
\frac{
b_{\s 1 \perp} + b_{\s 2 \perp} }{2}
\right|^2=\frac{\tau_{\s 2}^2\left|\overline{b}_{\s \perp}\right|^2}{2}.
\end{equation}
The meaning of Eq. (\ref{averaging}) is obvious: as a result of performing multiple ``short'' hops while awaiting  the ``long'' hop electron spin ``sees'' the
average hyperfine field, $\overline{b}_{\s \perp}$. If the number of sites visited in course of waiting was
big, the averaging of the corresponding hyperfine fields would lead to the suppression of $P_{\s \text{sf}}$.

We assumed that the net spin rotation is small, $b_{\s 0} \tau_{\s 2} \ll 1$.
 However, the above derivation suggests that we could impose a much weaker requirement, $b_{\s 0} \tau_{\s 1} \ll 1$. This is because the effective averaging
takes place over time $\sim \tau_{\s 1}$. If under the condition $\Omega \tau_{\s 1} \ll 1$ the product $b_{\s 0} \tau_{\s 2}$ is not small, then $P_{\s \text{sf}}$
is given by the full Eq. (\ref {sf1}) with ${\bm \Omega}$ replaced by the average
of the vectors ${\bm b}_{\s 1}$ and ${\bm b}_{\s 2}$.

\subsection{Bouncing in a strong external field}

In a strong external field, $B\gg b_{\s 0}$,
the net spin rotation is small both for weak,  $b_{\s 0}\tau_{\s 2} \ll 1$, and for strong,  $b_{\s 0}\tau_{\s 2} \gg 1$, hyperfine fields. In the limit $B\tau_{\s 2}\gg 1$, when electron spin
rotates many times around the external field while waiting for the hop $2\rightarrow R$, the waiting time drops out of $P_{\s \text{sf}}$, see Eq. (\ref{sf1}). Effect of bouncing on $P_{\s \text{sf}}$ can be studied perturbatively with
respect to hyperfine field. In a zeroth  order we have, $a_{\s 1}(t) = \exp\left( -\frac{iBt}{2} \right)$, $a_{\s 2}=0$. In the first order the expression for $a_{\s 2}(t)$ takes the form
\begin{equation}
\label{B-system}
a_{\s 2}(t) = -\frac{i}{2} \int\limits_0^t dt' \,
b_{\s \perp}(t')
\exp\left( - iBt' \right).
\end{equation}
It is convenient to subtract $\overline{b}_{\s \perp}$
from  $b_{\s \perp}(t')$ in the integrand and rewrite Eq. (\ref{B-system}) as
\begin{equation}
\label{dividing}
a_{\s 2}(t)=-\overline{b}_{\s \perp}\frac{1-\exp{(-iBt)}}{2B}+\tilde{a}_{\s 2}(t),
\end{equation}
where $\tilde{a}_{\s 2}(t)$  is determined by Eq. (\ref{B-system})
in which $b_{\s \perp}(t')$ in the integrand is replaced by the difference $\left(b_{\s \perp}(t')-\overline{b}_{\s \perp}\right)$.  The term  $\tilde{a}_{\s 2}(t)$ captures the effect of bouncing.
Next, it is convenient to divide the domain of integration
in  Eq. (\ref{B-system}) into $N=Bt/2\pi$
intervals $\frac{2\pi}{B}n<t^{\prime}<\frac{2\pi}{B}(n+1)$, and reduce
the integration to a single interval $0<t^{\prime}<\frac{2\pi}{B}$. This yields
\begin{equation}
\label{B-system1}
\tilde{a}_{\s 2}(t) = -\frac{i}{2}\sum_{n=0}^N \int\limits_0^{2\pi/B} dt' \,
\exp\left( - iBt' \right)\left[b_{\s \perp}\bigl(t'+\frac{2\pi}{B}n\bigr)-\overline{b}_{\s \perp}\right].
\end{equation}
In the domain $\frac{1}{\tau_{\s 2}}<B<\frac{1}{\tau_{\s 1}}$  the right-hand side of Eq. (\ref{B-system1}) is a sum of $N$ statistically-equivalent and independent terms, each being zero on average. In each term the integrand changes sign $2\pi/B\tau_{\s 1}$ times with magnitude $\Delta b_{\s \perp} = \frac{1}{2}|b_{\s 1\perp}-b_{\s 2\perp}|$. This allows us to estimate $|\tilde{a}_{\s 2}(t)|^2$ as follows
\begin{equation}
\label{B-system2}
|\tilde{a}_{\s 2}(t)|^2\sim |\Delta b_{\s \perp}|^2\tau_{\s 1}^2
\Biggl[N^{1/2}\Bigl(\frac{2\pi}{B\tau_{\s 1}}\Bigr)^{1/2}\Biggr]^2
\sim |\Delta b_{\s \perp}|^2t\tau_{\s 1}.
\end{equation}
The factor $N^{1/2}$ accounts for the fact that
all $N$ terms in the sum (\ref{B-system1}) are random.
The factor $\Bigl(\frac{2\pi}{B\tau_{\s 1}}\Bigr)$ accounts for the fact that {\em each} term is the sum
of $2\pi/B\tau_{\s 1}$ random contributions.

We see that magnetic field has dropped out of the ``bouncing''
estimate for $|\tilde{a}_{\s 2}(t)|^2$. It dominates over the
``regular'' part, given by the first term in Eq. (\ref{dividing}),
if $t\tau_{\s 1}\gg 1/B^2$. Since the characteristic $t$ is $\sim \tau_{\s 2}$, this defines a characteristic field
\begin{equation}
\label{characteristic}
B_c=\frac{1}{\left(\tau_{\s 1}\tau_{\s 2}\right)^{1/2}}.
\end{equation}
At $B \sim B_c$ the spin-flip probability crosses over from
$P_{\s \text{sf}}\sim b_{\s 0}^2/B^2$ to a plateau value
\begin{equation}
\label{plateau}
P_{\s \text{sf}}\sim b_{\s 0}^2 \tau_{\s 1} \tau_{\s 2}.
\end{equation}
In deriving Eq. (\ref{plateau}) we assumed that many bounces took place during the period, $2\pi/B$, of the in-plane spin rotation. This assumption is justified since $B_c\tau_{\s 1}
\sim \left(\tau_{\s 1}/\tau_{\s 2}\right)^{1/2} \ll 1$.
As magnetic field increases above $1/\tau_{\s 1}$,
the spin will execute many in-plane rotations in course of every bounce. Then the integral in the expression for
$\tilde{a}_{\s 2}$ can be viewed as a sum of
$\tau_{\s 2}/\tau_{\s 1}$ random contributions each of
being of the order of
$\Delta b_{\s \perp}/B$.
Then we can again estimate $\tilde{a}_{\s 2}$, and subsequently, $P_{\s \text{sf}}(B)$, from the variance. The result reads
\begin{equation}
\label{large-B}
P_{\s \text{sf}}(B)\Bigl|_{B\gg 1/\tau_{\s 1}}\sim \frac{|\Delta b_{\s \perp}|^2}{B^2}\left(\frac{\tau_{\s 2}}{\tau_{\s 1}}\right).
\end{equation}
Note that the bouncing-related spin-flip probability, Eq. (\ref{large-B}), exceeds the result $P_{\s \text{sf}}\sim
 b_{\s 0}^2/B^2$ in the absence of bouncing by a large
 factor $\frac{\tau_{\s 2}}{\tau_{\s 1}}$, which is the number of bounces.

Thus, unlike the case $B=0$, the bouncing causes the
growth of the spin-flip probability.
The probability  Eq. (\ref{large-B}) for strong fields matches $P_{\s \text{sf}}$ for intermediate  fields,
Eq. (\ref{plateau}), at $B\sim \frac{1}{\tau_{\s 1}}$.

\begin{figure}
\includegraphics[width=77mm,clip]{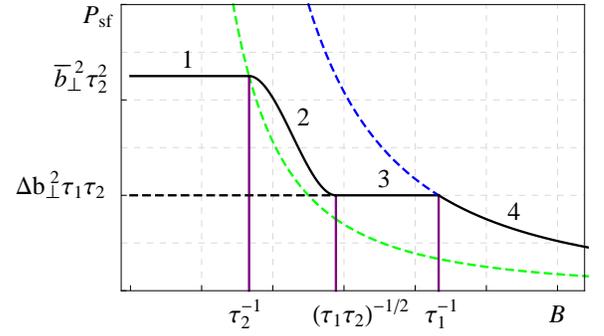}
\caption{(Color online) Schematic illustration of the enhancement of the spin-flip probability due to multiple bounces. In the absence of bouncing, the plateau $({\bm 1})$ at small external fields crosses over to the $1/B^2$ behavior (green dashed line) at $B\sim 1/\tau_{\s 2}$, where $\tau_{\s 2}$ is the waiting time   for the hop $2\rightarrow R$.
When the waiting time, $\tau_{\s 1}$, for the hops $1\rightarrow 2$ and $2\rightarrow 1$ is much
shorter than $\tau_{\s 2}$, the spin-flip probability decreases $({\bm 2})$, develops a second plateau $({\bm 3})$ at
$B \sim(\tau_{\s 1}\tau_{\s 2})^{-1/2}$, see Eq. (\ref{summarize}), and crosses over $({\bm 4})$ to
$1/B^2$ behavior (blue dashed line) at $B \sim 1/\tau_{\s 1}$.}
\label{fig-summary}
\end{figure}

In conclusion of the subsection we summarize the results
for $P_{\s \text{sf}}$ in different domains of magnetic field
\begin{equation}
\label{summarize}
P_{\s \text{sf}}(B)=\left\{ \begin{matrix}
\overline{b}_{\s \perp}^2 \tau_{\s 2}^2, & 0 < B < \frac{1}{\tau_{\s 2}} \\
(\Delta b_{\s \perp}\!)^2 \tau_{\s 1} \tau_{\s 2} + \frac{\overline{b}_{\s \perp}^2}{B^2},
& \frac{1}{\tau_{\s 2}} < B < \frac{1}{\tau_{\s 1}} \\
\frac{(\Delta b_{\s \perp}\!)^2}{B^2} \frac{\tau_{\s 2}}{\tau_{\s 1}}, & B> \frac{1}{\tau_{\s 2}}
\end{matrix}
\right.
\end{equation}
The evolution of the spin-flip probability with magnetic
field is illustrated in Fig. \ref{fig-summary}.

\section{Discussion}

\begin{itemize}
\item
Conventional treatments of spin relaxation neglect interference
effects.  This happens at the stage when the exact equation for the
density matrix is solved using the  ``tau-approximation'', see
e.g. the review Ref. \onlinecite{Glazov}. Concerning the effect of bouncing
considered in the  present paper, there is an analog of bouncing in
spin relaxation caused by spin-orbit coupling. In course of the {\em orbital} electron motion in a strong
magnetic field, it keeps returning to the origin after undergoing the same
sequence of scattering events. This ``memory'' results in shortening of the
spin relaxation time\cite{Glazov}. Similarly,  Eq. (\ref{probabilistic}), where bouncing is
treated probabilistically,  predicts that $P_{\s \text{sf}}$ approaches to $1/2$
faster as the number, $m$, of bounces grows.
We emphasize that the quantum treatment of bouncing leads to the opposite result.

\item Absence of the Hanle effect reported in Refs. \onlinecite{noHanle1} and \onlinecite{noHanle2}
can be interpreted as independence of $P_{\s \text{sf}}$ on the
{\em magnitude} of the external field.
In this regard, we note that the partial $p_{\s \text{sf}}$ values given
by Eq. (\ref{sf1}) increase monotonically with increasing magnetic field,
for any field orientation.
However, in Sect. II we demonstrated that, when the partial probabilities $p_{\s \text{sf}}$ are in the
vicinity of $1/2$, the dependence of the net two-step probability, $P_{\s \text{sf}}$,
on these partial probabilities is {\em non-monotonic}.
Moreover, the derivative of $P_{\s \text{sf}}$ with respect to the
magnetic field passes through zero.
This indicates that, for a range of parameters where $P_{\s \text{sf}}$
is near its maximum, there is no sensitivity to the magnitude of the
applied field.  Note however, that, since this behavior is a consequence
of interference, it does not survive averaging over hyperfine field
distributions.


\item
The fact that electron flips its spin as it travels between
the electrodes constitutes  an additional source of shot noise\cite{ZhenyaValve} and, thus,  affects the Fano factor.
The above calculation of $P_{\s \text{sf}}$
is insufficient to find the Fano factor of  ``two-site'' transport
with spin flip. The reason for this is that the transport of charge is
incoherent while the spin-transport is fully coherent.
Qualitatively, the complexity of description of noise
follows from the fact that the Fano factor must depend on {\em both}
$P_{\s \text{sf}}$ and the magnetizations
$\mathcal{P}_{\s 1}, \mathcal{P}_{\s 2}$ of the electrodes.
The latter conclusion can be inferred from the reasoning presented
in Ref. \onlinecite{ZhenyaValve}. Suppose that magnetizations of the electrodes are anti-parallel, and $P_{\s \text{sf}}$ is small. Then,
no matter what is the actual mechanism of transport, the Fano factor
should be $1$, which is the Poissonian value. This is because, in order
to be transferred between the electrodes, electron must flip the spin.
For small $P_{\s \text{sf}}$ it is  waiting time for the spin-flip which is the bottle-neck for transport, since it is much longer than
the waiting time for all hops.
All we can say is that, if  one
electrode changes from oppositely polarized  to non-polarized,
the Fano factor changes from $1$ to
$\frac{\tau_{\s 1}^2+\tau_{\s 1}\tau_{\s 2}+\tau_{\s 2}^2}{(\tau_{\s 1} + \tau_{\s 2})^2}$, which is the Fano factor for {\em spin-independent} transport
through the same sites.
Rigorous evaluation of the Fano factor with magnetization of electrodes
taken into account, requires solving the equation for time evolution of the full density matrix.

\item
As was mentioned in the Introduction, in diffusive transport, spin-memory
loss is incorporated via the ``survival'' probability, $\exp(-\text{L}/l_{\s s}),$ which we replaced by $1 - 2P_{\s \text{sf}}$.
The probabilistic description, on the other hand, predicts that $P_{\s \text{sf}}$ falls off exponentially with the number of
steps, $N$, see Eq. (\ref{NP_sf-prob}), or equivalently with time,
but {\em not} with length.
The exponential dependence $\exp(-\text{L}/l_{\s s})$ is recovered
upon the transformation
\begin{equation}
\label{transformation}
P_{\s \text{sf}}(\text{L}) = \int dN\, P_{\s \text{sf}}(N) \exp \left(
-\frac{\text{L}^2}{N r^2}
\right),
\end{equation}
where $r$ is the length of a diffusion step.  This yields
\begin{equation}
l_{\s s} = \frac{r}{\sqrt{|\ln(1-2p_{\s \text{sf}})|}}.
\end{equation}
Here we would like to emphasize that the concept of spin diffusion
length does not apply for multistep transport.\cite{Bobbert,multisteps}
The reason for this is twofold.
Unlike diffusion,
the relationship between $\text{L}$ and $r$ in multistep transport is
$r = \sqrt{\text{L} a}, $ and $N=\text{L}/r=\sqrt{\text{L}/a}$, where $a$ is the under-barrier
decay length.\cite{Tartakovskii}
Secondly, also unlike diffusion, the waiting time for the next
step, which is the time for spin precession, is also a function of $\text{L}$ and $N$, specifically, $\tau = \tau_{\s 0}\exp(2\text{L}/Na)$.  As a result we get
\begin{equation}
\label{multistep-chain}
P_{\s \text{sf}} = \frac{1}{2}\left[ \frac{  b_{\s 0}^2 \tau_{\s 0}^2 \exp\left( \sqrt{\frac{\text{L}}{a}} \right) }{
1+  b_{\s 0}^2 \tau_{\s 0}^2 \exp\left( \sqrt{\frac{\text{L}}{a}} \right) } \right].
\end{equation}
We see from Eq. (\ref{multistep-chain}) that for multistep transport
the spin-memory falls off with thickness of the active layer, $\text{L}$, slower than for diffusive transport. Anomalous dependence of TMR on the device thickness was reported   in Ref. \onlinecite{C60}. However Eq. (\ref{multistep-chain}) does not explain
the established facts that TMR vanished with increasing bias and temperature\cite{C60,Bader}.
\end{itemize}
\section{Acknowledgements}
We are grateful to Z. V. Vardeny for initiating this work. We also thank  E. G. Mishchenko for reading the manuscript, illuminating discussions, and suggestions on the presentation.
This work was supported by NSF through MRSEC DMR-1121252.

\begin{appendix}
\section{}
For spin-independent unidirectional transport the current between the electrodes can be viewed as a sequence of cycles
\begin{multline}
\label{nospin}
I(T) = \delta(T - T_{\s 1})
+ \delta(T - T_{\s 1} - T_{\s 2}) \\
+ \delta(T - T_{\s 1} - T_{\s 2} - T_{\s 3})
+ \cdots,
\end{multline}
where $T_{\s i}$ is a random  waiting time for the next electron to be transferred between $L$ and $R$. Suppose now that the left electrode is polarized $\uparrow$,
while the right electrode is polarized $\downarrow$. Then the electron transfer requires a spin-flip, and Eq. (\ref{nospin}) should be modified as
\begin{multline}
\label{withspin}
I_{\s \ud}(T) = P_{\s \text{sf}}(T_{\s 1})\delta(T - T_{\s 1})
+ P_{\s \text{sf}}(T_{\s 2})\delta(T - T_{\s 1} - T_{\s 2}) \\
+ P_{\s \text{sf}}(T_{\s 3})\delta(T - T_{\s 1} - T_{\s 2} - T_{\s 3})
+ \cdots,
\end{multline}
where $P_{\s \text{sf}}(T)$ is a quantum-mechanical probability that after
a composite process with duration $T$ the electron flips its spin. For the situation considered in this paper the composite process is an inelastic two-hop tunneling. To calculate the average current one should take the limit of large $T$ and average over the compound waiting times, $T_{\s i}$, with distribution function $F(T)$. This averaging is convenient to perform\cite{vonOppen} using the integral representation of the $\delta$-function. Then the sum Eq. (\ref{withspin}) turns into a geometrical progression, the summation of which yields
\begin{equation}
\langle I_{\s \ud}(T) \rangle
 =  \int \frac{d \alpha}{2 \pi} e^{i \alpha T}
 \frac{\langle P_{\s \text{sf}}(T') \exp\left( -i \alpha T' \right) \rangle}{1 - \langle \exp\left(-i \alpha T'\right) \rangle}.
\end{equation}
In the limit $T\rightarrow \infty$ one can set $\alpha=0$ in the numerator and expand the denominator to the lowest order. After that the integration over $\alpha$ can be easily performed leading to the natural result
\begin{equation}
\label{fraction}
\langle I_{\s \ud}\rangle
 = \frac{ \langle P_{\s \text{sf}} \rangle }{ \langle T \rangle },
\end{equation}
where $\langle P_{\s \text{sf}}\rangle$ is defined as
\begin{equation}
\label{T}
\langle P_{\s \text{sf}}\rangle = \int P_{\s \text{sf}}(T) F(T) dT.
\end{equation}
For a particular case of a two-hop transport we have $T=t_{\s 1}+t_{\s 2}$, where
$t_{\s 1}$ and $t_2$ are distributed with $f_{\s 1,2}(t)=\frac{1}{\tau_{\s 1,2}}\exp(-t/\tau_{\s 1,2})$. Then Eq. (\ref{T}) assumes the form
\begin{equation}
\langle P_{\s \text{sf}}\rangle =  \int_{0}^{\infty} dt_{\s 1} \int_{0}^{\infty} dt_{\s 2} \;
P_{\s \text{sf}}(t_{\s 1}, t_{\s 2})  f_{\s 1}(t_{\s 1}) f_{\s 2}(t_{\s 2}).
\end{equation}
This is exactly the quantity calculated in Sect. II. From Eq. (\ref{fraction}) we
conclude that for calculation of average current one should multiply this quantity by $1/\langle T \rangle$, which is the current between unpolarized electrodes.

From the same reasoning we confirm that opposite directions of polarization of the electrodes the current is equal to $I_{\s \uu}=(1-\langle P_{\s \text{sf}}\rangle)/\langle T \rangle$. Therefore the expression for TMR with completely polarized electrodes takes the form
\begin{equation}
\text{TMR} = \frac{I_{\s \uu} - I_{\s \ud}}{I_{\s \uu} + I_{\s \ud}}
 = 1 - 2 P_{\s \text{sf}}.
\end{equation}

For partial polarization of electrodes with concentrations $\mathcal{N}_{\s \uparrow}$, $\mathcal{N}_{\s \downarrow}$ of $\uparrow$ and $\downarrow$ electrons in the left electrode and $n_{\s \uparrow}$, $n_{\s \uparrow}$ in
the right electrode, the general expressions for $I_{\s \uu}$ and $I_{\s \ud}$ can be presented as
\begin{align}
\label{uu}
I_{\s \uu} &= \Gamma_{\s \uu} \left( \mathcal{N}_{\s \uparrow} n_{\s \uparrow}  + \mathcal{N}_{\s \downarrow} n_{\s \downarrow} \right)
+ \Gamma_{\s \ud} \left( \mathcal{N}_{\s \uparrow} n_{\s \downarrow}  + \mathcal{N}_{\s \downarrow} n_{\s \uparrow} \right),\\ \label{ud}
I_{\s \ud} &=
\Gamma_{\s \uu}  \left( \mathcal{N}_{\s \uparrow} n_{\s \downarrow}  + \mathcal{N}_{\s \downarrow} n_{\s \uparrow} \right)
+ \Gamma_{\s \ud}  \left( \mathcal{N}_{\s \uparrow} n_{\s \uparrow}  + \mathcal{N}_{\s \downarrow} n_{\s \downarrow} \right),
\end{align}
where $\Gamma_{\s \uu}$ and $\Gamma_{\s \ud}$ are the rates for the transfer processes from $\uparrow$ to $\uparrow$ and from $\uparrow$ to $\downarrow$. These rates are the characteristics of the active layer and do not depend on the polarizations of the electrodes. Naturally, we have  $\Gamma_{\s \uu}=\Gamma_{\s \dd}$ and $\Gamma_{\s \ud}=\Gamma_{\s \du}$. The expression Eq. (\ref{TMR1}) follows from Eqs. (\ref{uu}) and (\ref{ud}) in two steps. We relate the concentration via the degrees of polarization as
\begin{equation}
\mathcal{P}_{\s 1} = \frac{\mathcal{N}_{\s \uparrow} - \mathcal{N}_{\s \downarrow}}{\mathcal{N}_{\s \uparrow} + \mathcal{N}_{\s \downarrow}},\quad
\mathcal{P}_{\s 2} = \frac{n_{\s \uparrow} - n_{\s \downarrow}}{n_{\s \uparrow} + n_{\s \downarrow}},
\end{equation}
yielding
\begin{equation}
\text{TMR} = \frac{2 \mathcal{P}_{\s 1}\mathcal{P}_{\s 2} \left( \Gamma_{\s \uu} - \Gamma_{\s \ud}\right)}{
\left( \Gamma_{\s \uu} + \Gamma_{\s \ud} \right)
- \mathcal{P}_{\s 1}\mathcal{P}_{\s 2}\left( \Gamma_{\s \uu} - \Gamma_{\s \ud}\right)
}.
\end{equation}
Finally, we relate the rates $\Gamma_{\s \uu}$, $\Gamma_{\s \ud}$  via
$P_{\s \text{sf}}$ as
\begin{equation}
\frac{\Gamma_{\s \uu} - \Gamma_{\s \ud}}{
 \Gamma_{\s \uu} + \Gamma_{\s \ud} }
 = 1 - 2 P_{\s \text{sf}},
\end{equation}
and arrive to Eq. (\ref{TMR1}).

\end{appendix}

\end{document}